\documentclass[11pt]{article}
\usepackage{moriond}
\usepackage{lineno}

\def\mco{\multicolumn}

\def\be{\begin{equation}}
\def\ee{\end{equation}}
\def\bea{\begin{eqnarray}}
\def\eea{\end{eqnarray}}

\newcommand{\Photo}{}

\begin{document}
\vspace*{4cm}
\title{TOP QUARK PRODUCTION AT THE LHC}

\author{ M. AOKI\\
on behalf of the ATLAS and CMS Collaborations}

\address{Kobayashi-Maskawa Institute for the Origin of Particles and the Universe (KMI),\\ 
Nagoya University, Furo-cho, Chikusa-ku, Nagoya, Aichi  464-8602, Japan}

\maketitle\abstracts{
Measurements of the production cross section for top quark pairs and the single top quark by the ATLAS and CMS collaborations at the LHC are presented. Improved measurements at $\sqrt{s}=$7 TeV  as well as new measurements at $\sqrt{s}=$8 TeV are highlighted. All the measurements are in good agreement with the standard model predictions.
}

\section{Introduction}
The top quark is the heaviest known elementary particle, whose mass is close to the scale of the electroweak symmetry breaking.
If there is physics beyond the standard model in this field, it may appear with the top quark.
Events with one or two top quarks are also expected to be a large background for several new physics searches.
Precise determination of the production and the kinematics of the top quark would increase the sensitivity of such searches.

\section{Production cross section of the top quark pair}

\begin{table}[b]
\begin{footnotesize}
\caption{A summary of the recent measurements of the inclusive top quark pair production cross section~\protect\cite{bib2,bib3,bib4,bib5,bib6,bib7,bib8}.}
\label{tbl:sigma_ttbar_inclusive}
\vspace{0.4cm}
\begin{center}
\begin{tabular}{|c|c|c|c|c|c|c|c|c|l|}
\hline
Mode & Experiment & $\sqrt{s}$~[TeV] & $\int{\cal L}dt$~[${\rm fb}^{-1}$] & $\sigma_{t\bar{t}}$~[pb]  & $\Delta\sigma({\rm stat})$ &  $\Delta\sigma({\rm syst})$ &  $\Delta\sigma({\cal L})$ \\
\hline
Dilepton~\cite{bib2} & CMS & 7 & 2.3  & 161.9  & $\pm2.5$  & $^{+5.1}_{-5.0}$  & $\pm3.6$  \\
\hline 
$\ell$+jets with soft $\mu$ tag~\cite{bib3} & ATLAS & 7 & 4.7  & 165  & $\pm2$  & $\pm17$  & $\pm3$ \\
\hline 
$\tau_{\rm had}$+jets~\cite{bib4} & ATLAS & 7 & 1.7  & 194  & $\pm18$  & \mco{2}{|c|}{$\pm46$ } \\
\hline 
$\tau_{\rm had}$+jets~\cite{bib5} & CMS & 7 & 3.9  & 152  & $\pm12$  & $\pm32$  & $\pm3$ \\
\hline 
\hline
$\ell$+jets~\cite{bib6} & ATLAS & 8 & 5.8  & 241  & $\pm2$  & $\pm31$  & $\pm9$ \\
\hline 
$\ell$+jets~\cite{bib7} & CMS & 8 & 2.8  & 228  & $\pm9$  & $^{+29}_{-26}$  & $\pm10$ \\
\hline 
Dilepton~\cite{bib8} & CMS & 8 & 2.4  & 227  & $\pm3$  & $\pm11$  & $\pm10$  \\
\hline 
\end{tabular}
\end{center}
\end{footnotesize}
\end{table}

In the standard model, the production rate of the top quark pair is expected to be $164^{+13}_{-10}~{\rm pb}$ at $\sqrt{s}=7~{\rm TeV}$, and $238^{+22}_{-24}~{\rm pb}$ at $\sqrt{s}=8~{\rm TeV}$~\cite{theo_ttbar}.
This rate has been measured by the ATLAS~\cite{atlas} and CMS~\cite{cms} collaborations with high statistics data samples.
The recent result of the LHC-combined inclusive top quark pair production cross section at $\sqrt{s}=7~{\rm TeV}$ is $173.3\pm2.3({\rm stat})\pm9.8({\rm syst})~{\rm pb}$~\cite{bib1}.
Further improvements and understanding of $t\bar{t}$ production have been made by both collaborations, which are summarized in Table~\ref{tbl:sigma_ttbar_inclusive}.
The irreducible systematic uncertainty, which is coming from the signal modeling, is taking a large part in most of the measurements, therefore, it is important to measure the differential cross section in order to reduce this uncertainty.

%

\begin{figure}
\centerline{
\includegraphics[width=0.30\linewidth]{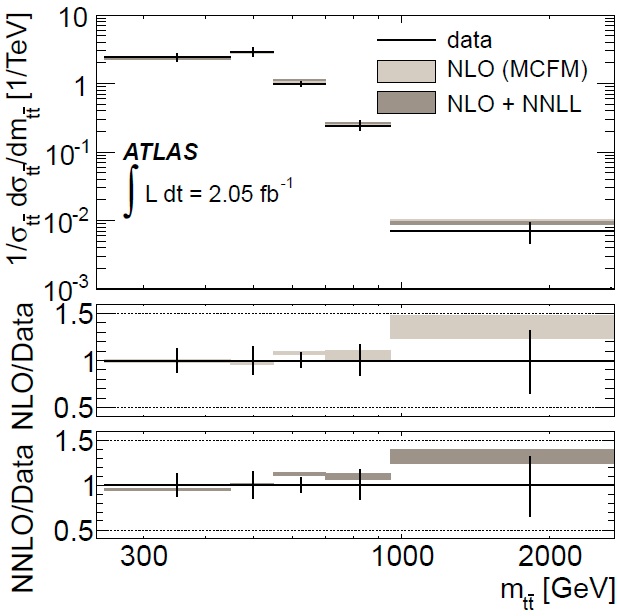}
\includegraphics[width=0.31\linewidth]{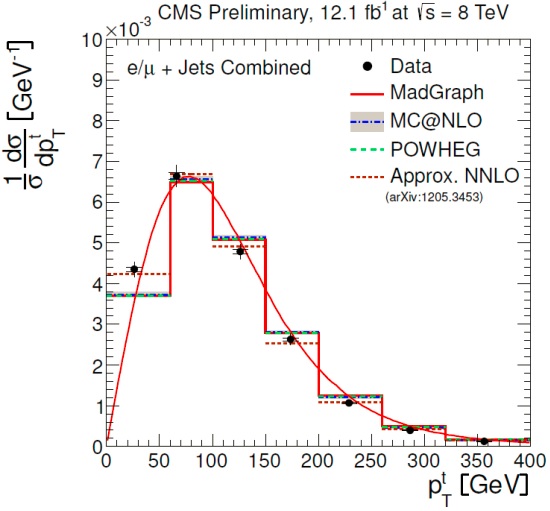}
\includegraphics[width=0.31\linewidth]{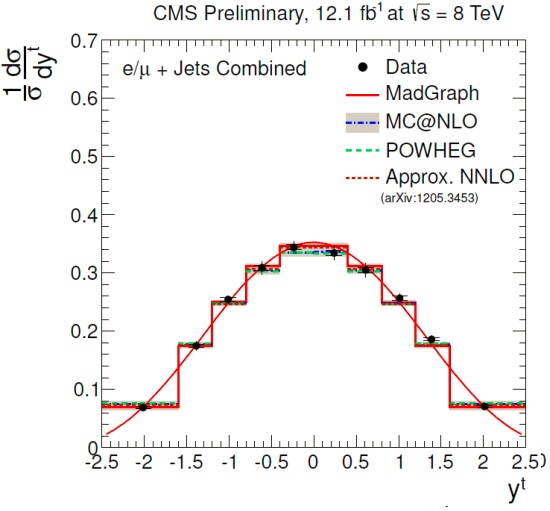}
}
\caption{Differential cross section as a function of $m_{t\bar{t}}$, $p_{T}^t$ and $y^t$~\protect\cite{bib9}.}
\label{fig:sigma_ttbar_differential}
\end{figure}

In Figure~\ref{fig:sigma_ttbar_differential}, the normalized differential cross sections as a function of $t\bar{t}$ invariant mass ($m_{t\bar{t}}$), transverse momentum of a top quark ($p_{T}^t$), and rapidity of a top quark ($y^t$) are shown~\cite{bib9}.  
The measured distributions agree well with the theoretical predictions within experimental uncertainties, and the approximate-NNLO provides the best description of the data.

%
%

\begin{figure}
\centerline{
\includegraphics[width=0.30\linewidth]{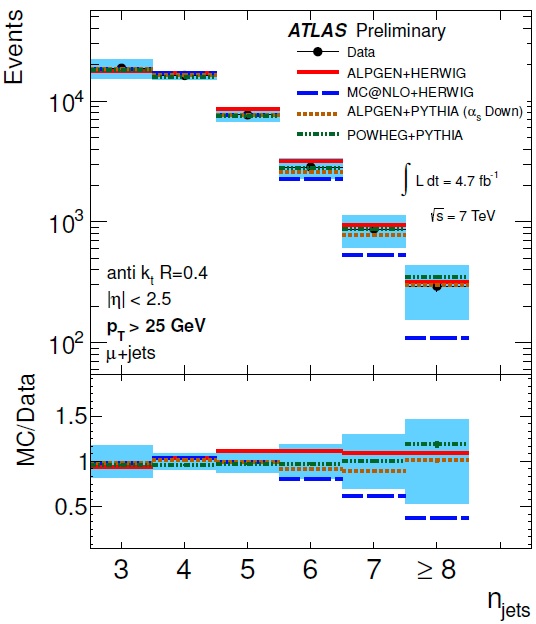}
\includegraphics[width=0.43\linewidth]{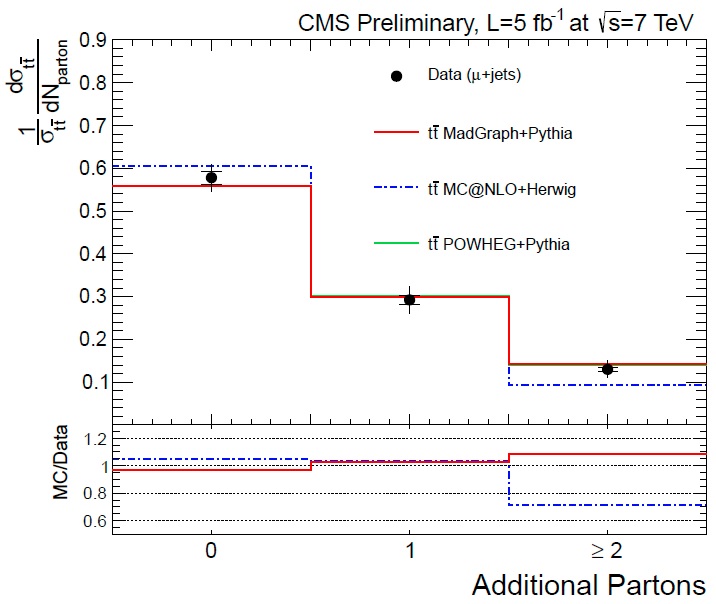}
}
\caption{Distributions of the jet multiplicity in $t\bar{t}$ events. Left: inclusive jet multiplicity. Right:jet multiplicity subtracting jets from $t\bar{t}$~\protect\cite{bib13}. }
\label{fig:additional_jets}
\end{figure}

Measurements of jets associated with the $t\bar{t}$ production could constrain models of initial and final state radiation at the scale of the top quark mass, and also provide a test of perturbative QCD in the LHC energy regime.
Figure~\ref{fig:additional_jets} shows the distributions of the inclusive jet multiplicity in $t\bar{t}$ events, and the distributions of the jet multiplicity subtracting those from the $t\bar{t}$ decay~\cite{bib13}. 
All the theoretical predictions agree well with the data, although the MC@NLO+HERWIG
prediction tends to underestimate the data for events in the higher jet multiplicity bins.

\begin{figure}
\centerline{
\includegraphics[width=0.33\linewidth]{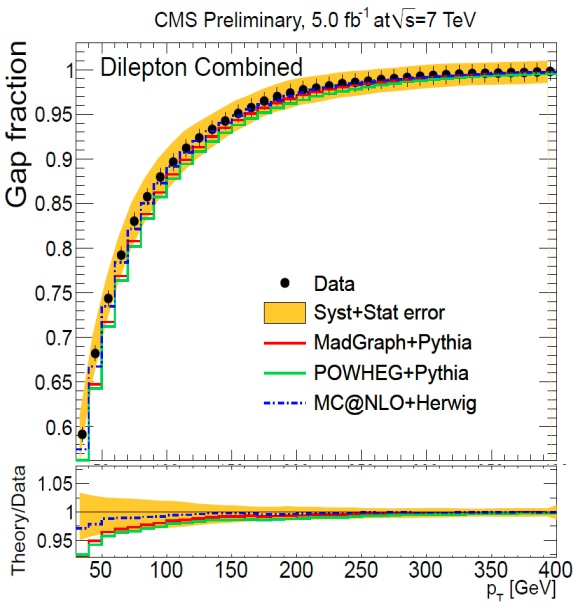}
\includegraphics[width=0.33\linewidth]{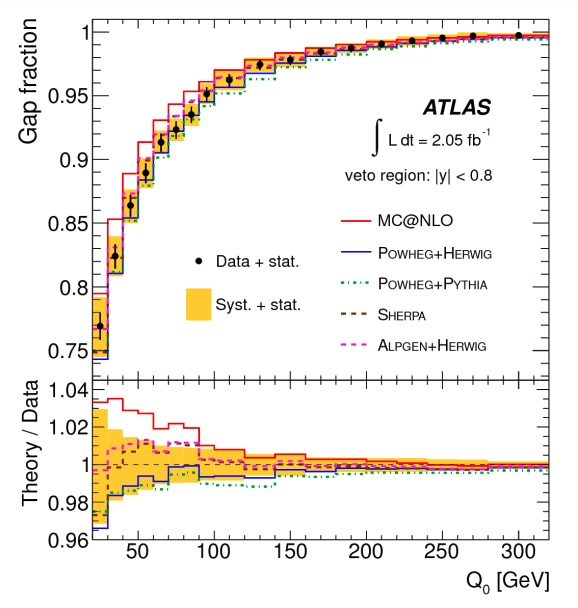}
\includegraphics[width=0.33\linewidth]{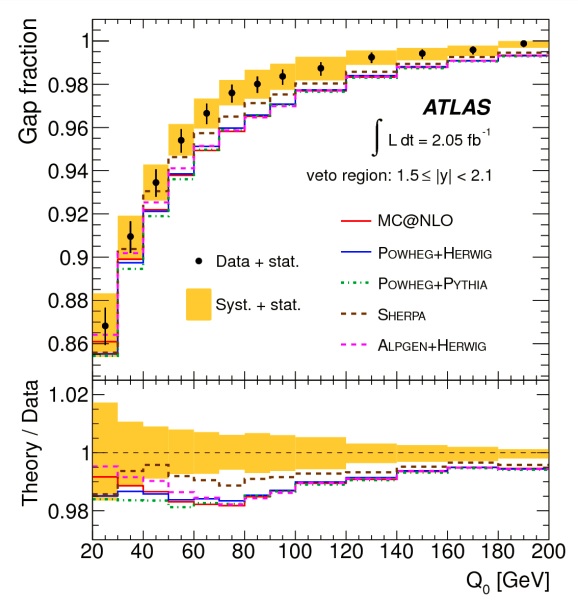}
}
\caption{Distributions of the gap fraction. Left: overall detector acceptance. Center:central rapidity region. Right: forward rapidity region~\protect\cite{bib15}.}
\label{fig:gap_fraction}
\end{figure}

An alternate method to see the jet activity is to measure the gap fraction, 
which is defined as the fraction of events with no additional jets with transverse momentum above a threshold.
Figure~\ref{fig:gap_fraction} shows distributions of the gap fraction measured in the dilepton channel~\cite{bib15}.
All the theoretical predications agree well with the data within experimental uncertainty for the overall rapidity region.
However, the MC@NLO prediction tends to have higher gap faction in the central rapidity region, and none of the predictions agrees with data in the forward rapidity region.

Another interest is to measure the flavor of such jets.
The CMS collaboration has performed the first measurement of the cross section ratio $\sigma(pp \to t\bar{t}b\bar{b}) / \sigma(pp \to t\bar{t}jj)$, where the $t\bar{t}jj$ final state is defined by the presence of two additional jets in addition to the two $b$ quarks from the decays of the top quark pair.
The resulting measured cross section ratio is $[ 3.6\pm1.1({\rm stat})\pm0.9({\rm syst}) ]\%$~\cite{bib17}.
This number can be compared to
the predictions using MADGRAPH~\cite{madgraph}
and POWHEG~\cite{powheg}, which are 1.2\% and 1.3\%, respectively

\section{Single top production}
Single top production is an alternative top quark production mechanism via the weak interaction.
Measurement of this production rate allows for a measurement of $|V_{tb}|$ without assumptions about the number of quark generations.
Three subprocesses contribute to single top quark production: 
the exchange of a virtual $W$ boson in the $t$-channel, or in the $s$-channel, and the associated production
of a top quark and an on-shell $W$ boson.

The ATLAS collaboration has reported evidence of the associated production of a top quark and a $W$ boson using $2.05~{\rm fb}^{-1}$ of 7 TeV data.
The signal is extracted using a multivariate analysis technique in the analysis.
The measured production cross section is $16.8\pm2.9({\rm stat})\pm4.9({\rm syst})~{\rm pb}$ with a significance of 3.3 standard deviations, and $|V_{tb}|$ is determined as  $1.03^{+0.16}_{-0.19}$~\cite{bib18}.
This is also confirmed by the CMS collaboration with more statistics, $4.9~{\rm fb}^{-1}$ of 7 TeV data. 
The reported production cross section is $16^{+5}_{-4}~{\rm pb}$ with a significance of 4.0 standard deviations, and $|V_{tb}|$ is measured to be $1.01^{+0.16}_{-0.13}({\rm exp})^{+0.03}_{-0.04}({\rm th})$~\cite{bib19}.
In the standard model, this production cross section is expected to be $15.7\pm1.1~{\rm pb}$~\cite{theo_wt}.

With a higher beam energy of $\sqrt{s}=$8 TeV, the single top production in the $t$-channel has been confirmed by both the ATLAS and the CMS collaborations.
In the standard model, this production cross section is expected to be $87.8^{+3.4}_{-1.9}~{\rm pb}$~\cite{theo_tch}
The signal is extracted, for ATLAS, by a fit to the output distribution of a neural network discriminant in $5.8~{\rm fb}^{-1}$ of data.
The measured production cross section is $95.1 \pm 2.4({\rm stat}) \pm 18.0({\rm syst})~{\rm pb}$, and $|V_{tb}|$ is measured to be $1.04^{+0.10}_{-0.11}$~\cite{bib20}.
The CMS collaboration measures this production using $5.0~{\rm fb}^{-1}$ of data.
The signal is extracted by a fit to the pseudo-rapidity distribution of the recoil jet,
and the production cross section is measured to be $80.1\pm5.7({\rm stat})\pm11.0({\rm syst})\pm4.0({\rm lumi})~{\rm pb}$, and resulting $|V_{tb}|$ is found to be $0.96\pm0.08({\rm exp})\pm0.02({\rm th})$~\cite{bib21}. 

\begin{figure}
\centerline{
\includegraphics[width=0.49\linewidth]{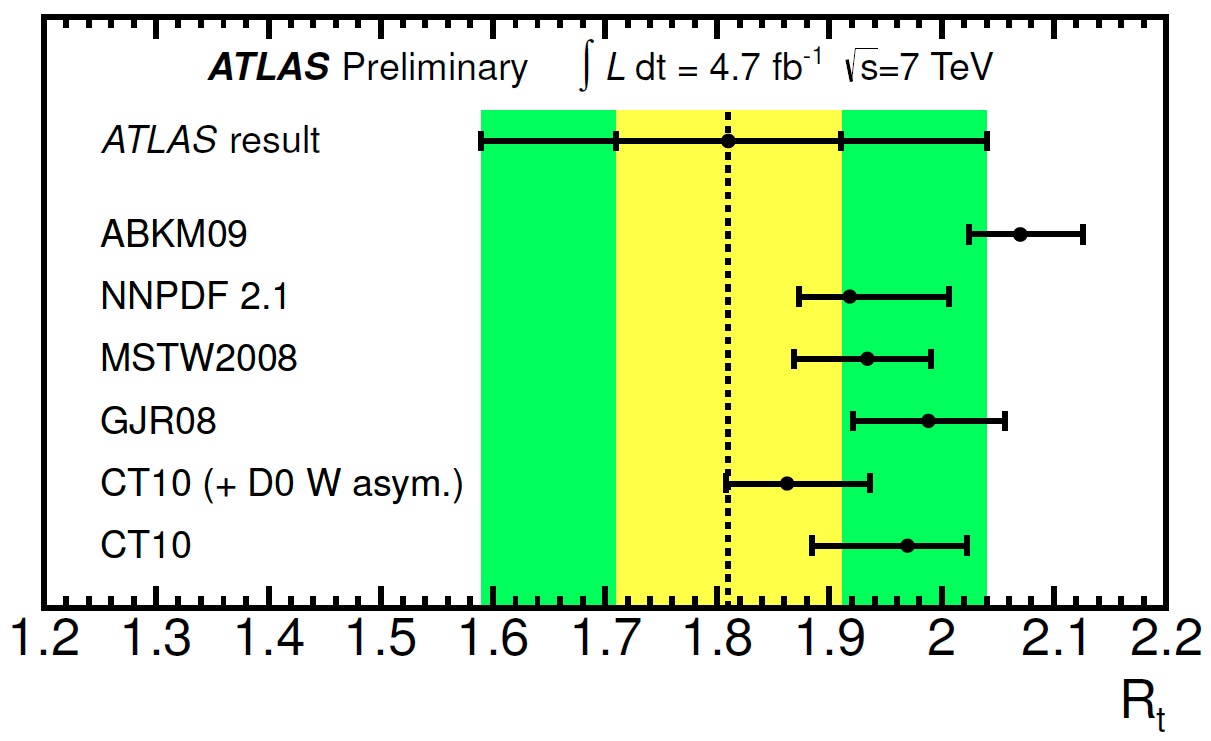}
\includegraphics[width=0.41\linewidth]{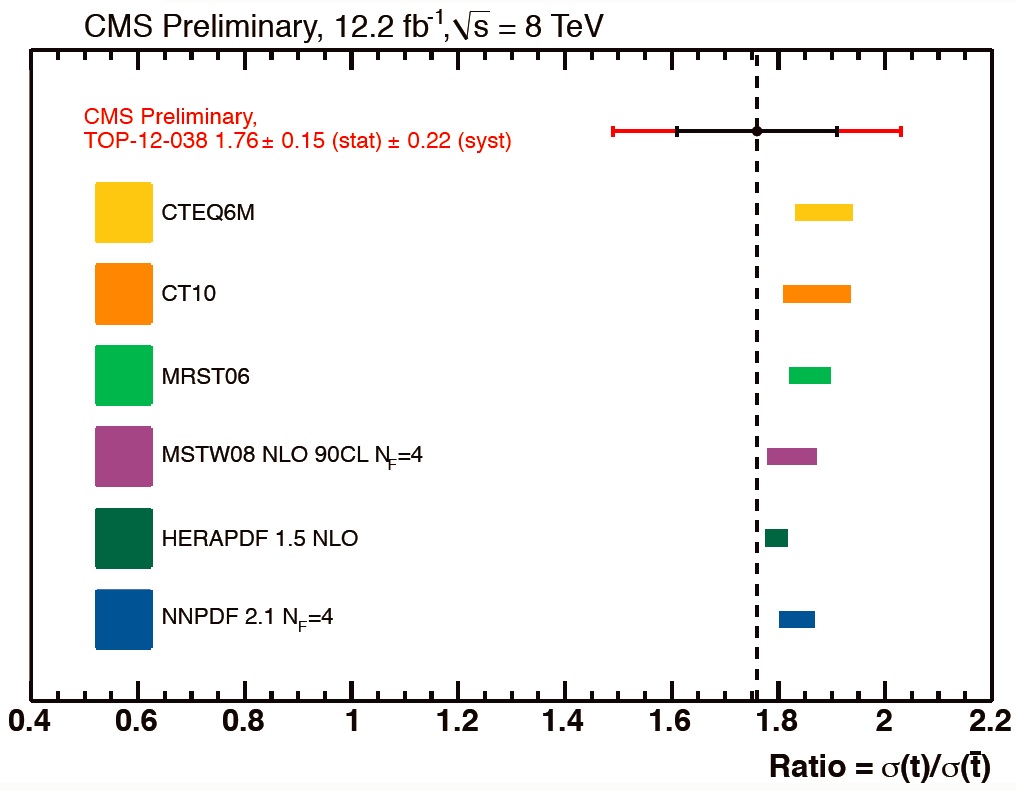}
}
\caption{$R_t$ results compared to several theoretical predictions~\protect\cite{bib22,bib23}.}
\label{fig:rt}
\end{figure}

The cross-section ratio between top and anti-top quarks, $R_t=\sigma_t/\sigma_{\bar{t}}$, in single top quark production is sensitive to the momentum probability density of quarks in the proton.
The up-quark density inside the proton is about twice as high as the down-quark density, and the production cross-section of single top-quarks is about twice as high as the cross section for anti-top quark production.
This cross section ratio has been measured by the ATLAS collaboration using $4.7~{\rm fb}^{_-1}$ of 7 TeV data,
and they report that $R_t$ is $1.81\pm0.10({\rm stat})^{+0.21}_{-0.20}({\rm syst})$~\cite{bib22}.
This ratio has been also measured by the CMS collaboration using $12.2~{\rm fb}^{-1}$ of 8 TeV data, and $R_t$ has been determined as $1.76\pm0.15({\rm stat})\pm0.22({\rm syst})$~\cite{bib23}. 
In Fig.~\ref{fig:rt}, the measured $R_t$ is compared with expectation using several different PDF sets.

\section{Conclusions}
Measurements of top quark production by the ATLAS and CMS collaborations have been presented. 
All those results are in good agreement with the standard model predictions.
The high luminosity of the LHC proton-proton collider at a TeV energy allows for high precision and detailed  measurements of top quark production.

\end{document}